\documentclass[fleqn,usenatbib]{mnras}

\usepackage[T1]{fontenc}
\usepackage{ae,aecompl}

\usepackage{newtxtext,newtxmath}

\usepackage[usenames,dvipsnames]{color}
\usepackage{graphicx}	
\usepackage{amsmath}	
\usepackage{amssymb}	
\usepackage{xspace}		
\usepackage{multicol}   
\usepackage[detect-all]{siunitx}  
\DeclareSIUnit\percentage{\,per\>cent}  
\DeclareSIUnit\erg{erg}  
\DeclareSIUnit\solarradius{\ensuremath{\text{R}_\odot}}
\DeclareSIUnit\solarmass{\ensuremath{\text{M}_\odot}}
\DeclareSIUnit\solarlum{\ensuremath{\text{L}_\odot}}
\DeclareSIUnit\angstrom{\ensuremath{\textup{\AA}}}

\newcommand{\mystar}{HR~7322\xspace}
\newcommand{\ldc}{\ensuremath{u_{\lambda}}\xspace}
\newcommand{\numax}{\ensuremath{\nu_\text{max}}\xspace}
\newcommand{\dnu}{\ensuremath{\Delta\nu}\xspace}
\newcommand{\fdnu}{\ensuremath{f_{\dnu}}\xspace}
\newcommand{\fnumax}{\ensuremath{f_{{\nu}_{\text{max}}}}\xspace}
\newcommand{\angdia}{\ensuremath{\theta_{\mathrm{LD}}}\xspace}
\newcommand{\teff}{\ensuremath{T_{\textup{eff}}}\xspace}
\newcommand{\feh}{\ensuremath{[\text{Fe/H}]}\xspace}
\newcommand{\meh}{\ensuremath{[\text{Me/H}]}\xspace}
\newcommand{\alphafe}{[\ensuremath{\text{$\alpha$/Fe}]}\xspace}
\newcommand{\amlt}{\ensuremath{\alpha_{\textup{mlt}}}\xspace}
\newcommand{\fbol}{\ensuremath{F_{\textup{bol}}}\xspace}

\newcommand{\stagger}{\textsc{stagger}\xspace}
\newcommand{\garstec}{\textsc{garstec}\xspace}
\newcommand{\adipls}{\textsc{adipls}\xspace}
\newcommand{\basta}{\textsc{basta}\xspace}
\newcommand{\starradius}{\ensuremath{R}\xspace}
\newcommand{\starmass}{\ensuremath{M}\xspace}

\title[The subgiant HR 7322 as a benchmark for asteroseismology]{The subgiant HR 7322 as an asteroseismic benchmark star}

\author[Stokholm et al.]{Amalie Stokholm,$^{1}$\thanks{E-mail: stokholm@phys.au.dk} 
Poul Erik Nissen,$^{1}$ 
V\'ictor Silva Aguirre,$^{1}$
Timothy R. White,$^{1,2}$\newauthor
Mikkel N. Lund,$^{1,3}$, Jakob R\o rsted Mosumgaard$^{1}$, Daniel Huber$^{4,5,6,1}$, and\newauthor Jens Jessen-Hansen$^{1}$
\\
$^{1}$Stellar Astrophysics Centre, Department of Physics and Astronomy, Aarhus University, Ny~Munkegade~120, DK-8000 Aarhus C, Denmark.\\
$^{2}$Research School of Astronomy and Astrophysics, Australian National University, Canberra, ACT 2611, Australia \\
$^{3}$School of Physics \& Astronomy, University of Birmingham, Edgbaston, Birmingham, B15 2TT, UK \\
$^{4}$Institute for Astronomy, University of Hawai`i, 2680 Woodlawn Drive, Honolulu, HI 96822, USA \\
$^{5}$Sydney Institute for Astronomy (SIfA), School of Physics, University of Sydney, NSW 2006, Australia \\
$^{6}$SETI Institute, 189 Bernardo Avenue, Mountain View, CA 94043, USA
}

\date{Accepted XXX. Received YYY; in original form ZZZ}

\pubyear{2019}

\begin{document}
\label{firstpage}
\pagerange{\pageref{firstpage}--\pageref{lastpage}}
\maketitle

\begin{abstract}
We present an in-depth analysis of the bright subgiant HR~7322 (KIC~10005473) using {\it Kepler} short-cadence photometry, optical interferometry from CHARA, high-resolution spectra from SONG, and stellar modelling using {\textsc{garstec}} grids and the Bayesian grid-fitting algorithm {\textsc{basta}}. HR~7322 is only the second subgiant with high-quality {\it Kepler} asteroseismology for which we also have interferometric data.
We find a limb-darkened angular diameter of $0.443 \pm 0.007$~mas, which, combined with a distance derived using the parallax from {\it Gaia} DR2 and a bolometric flux, yields a linear radius of $2.00 \pm 0.03$~R$_{\odot}$ and an effective temperature of $6350 \pm 90$~K. HR~7322 exhibits solar-like oscillations, and using the asteroseismic scaling relations and revisions thereof, we find good agreement between asteroseismic and interferometric stellar radius. The level of precision reached by the careful modelling is to a great extent due to the presence of an avoided crossing in the dipole oscillation mode pattern of HR~7322.
We find that the standard models predict a stellar radius systematically smaller than the observed interferometric one and that a sub-solar mixing length parameter is needed to achieve a good fit to individual oscillation frequencies, interferometric temperature, and spectroscopic metallicity.
\end{abstract}




\section{Introduction}
\label{sec:introduction}
Understanding the evolution and structure of stars is one of the key challenges in modern astrophysics. One way to unravel the secrets of stellar interiors is to compare models of stellar structure and evolution with precise observations of stars of different masses and evolutionary stages.
Our ability to test and improve stellar models thus rely crucially on the information available to constrain the parameter space and our understanding of the stars is therefore driven by advances in measuring stellar properties precisely and accurately.

One method to precisely determine stellar parameters is asteroseismology, the study of stellar oscillations. The turbulent, convective motion beneath the stellar photosphere of stars like the Sun stochastically excite and damp acoustic waves within the star and these stellar pulsations extend through the otherwise opaque stellar interior \citep{goldreich1977}. The stellar oscillations can be measured e.g.\@ by observing how the brightness of the star subtly vary as a function of time. If a pulsating star is observed for a sufficiently long period of time \citep[approximately ten times the mode lifetime,][]{garcia2015}, the individual mode oscillations can be extracted from the Fourier transform of the time series, the so-called power spectrum \citep[see e.g.\@][and references therein]{appourchaux2013,garcia2015,campante2018}.  For solar-type stars on the main sequence, the observed modes are acoustic modes or p-modes for which the restoring force of the oscillating motion arises from the pressure gradient. The stellar oscillations are sensitive to the conditions in the stellar interior and thus studying the frequency pattern of the different pulsation modes reveals information about the internal structure and composition of the star \citep[see e.g.\@][]{brown1994,aerts2010,chaplin2013,verma2014,deheuvels2016,basu2017,hekker2017}.

Scaling relations of the stellar radius (\starradius) and mass (\starmass) can be derived from two global asteroseismic parameters of the oscillation pattern along with an estimate of effective temperature (\teff) by scaling the value of the Sun to the observed quantities.
The power of the observed modes has an envelope with a Gaussian-like shape and thus one of these global asteroseismic parameters is the frequency of maximum power $\numax$ and, as it is expected to scale with the acoustic cut-off frequency \citep{brown1991,kjeldsen1995}, is related to global stellar properties through a semi-empirical scaling relation of the form 
\begin{equation}
\label{eq:numaxscalingrelation}
\frac{\numax}{\nu_{\text{max,}\odot}} \simeq \left(\frac{\starmass}{\text{M}_{\odot}}\right) \left(\frac{\starradius}{\text{R}_{\odot}}\right)^{-2} \left(\frac{\teff}{\text{T}_{\text{eff,}\odot}}\right)^{-1/2}.
\end{equation}
The observed p-modes are known to be approximately regularly spaced in frequency \citep{tassoul1980,scherrer1983}, and therefore the other global asteroseismic parameter is the large frequency separation $\dnu$, defined as the average separation in frequency between consecutive radial overtones $n$ of the same spherical degree $l$. The square of the large frequency separation can be shown analytically to be related to the mean density of the star \citep{ulrich1986} and thus another widely-used scaling relation is
\begin{equation}
\label{eq:dnuscalingrelation}
\frac{\dnu}{\Delta\nu_{\odot}} \simeq \left(\frac{\starmass}{\text{M}_{\odot}}\right)^{1/2} \left(\frac{\starradius}{\text{R}_{\odot}}\right)^{-3/2}.
\end{equation}
If Eqs.\@\ref{eq:numaxscalingrelation}~and~\ref{eq:dnuscalingrelation} are solved for mass and radius, we find that
	\begin{equation}
	\label{eq:mscalingrelation}
	\frac{\starmass}{\text{M}_{\odot}} \simeq \left(\frac{\numax}{\nu_{\text{max,}\odot}}\right)^{3} \left(\frac{\dnu}{\Delta\nu_{\odot}}\right)^{-4} \left(\frac{\teff}{\text{T}_{\text{eff,}\odot}}\right)^{3/2},
	\end{equation}
and
	\begin{equation}
	\label{eq:rscalingrelation}
	\frac{\starradius}{\text{R}_{\odot}} \simeq \left(\frac{\numax}{\nu_{\text{max,}\odot}}\right) \left(\frac{\dnu}{\Delta\nu_{\odot}}\right)^{-2} \left(\frac{\teff}{\text{T}_{\text{eff,}\odot}}\right)^{1/2}.
	\end{equation}
Using Eqs.~\ref{eq:mscalingrelation}~and~\ref{eq:rscalingrelation} to estimate the stellar mass and radius is often referred to as the \emph{direct method} \citep[e.g.][]{silvaaguirre2012,huber2017}. As these scaling relations are extrapolated from the Sun, the validity of scaling relations as a function of evolutionary state, metallicity, and effective temperature is currently an active topic within the field. 

If we want to test the validity of the asteroseismic scaling relations, we need to compare the radii and masses obtained from the direct method to other measurements. Recently, \citet{huber2017} tested the validity of the asteroseismic scaling relations on a sample of 2200 stars of different evolutionary state by comparing their asteroseismic radii to radii extracted from \emph{Gaia} DR1 (TGAS) \citep{gaia2016dr1,lindegren2016} and found that the asteroseismic radii were accurate to ${\sim}\SI{5}{\percentage}$ or better for stars with radii between $0.8$--$\SI{8}{\solarradius}$. However, they found that the radii of the subgiant stars seem to be systematically underestimated compared to radii from \emph{Gaia} and presents a larger scatter than the rest of their sample. This systematic offset hints at the need for good benchmark stars in this evolutionary state.

A different method of precisely measuring stellar parameters is long-baseline optical interferometry, an observational method in which the interference of light is used to obtain great angular resolution. The contrast between the dark and bright patches in the interference pattern (known as the visibility) at a given wavelength and baseline is directly related to the angular size of the observed object \citep{vancittert1934, zernike1938}.
Using trigonometry, we see that combining the angular size $\theta$ of a star, measured in radians, with a distance~($D$) yields the linear radius of the star,
	\begin{equation}
	\label{eq:linearradius}
	\starradius = \frac{1}{2}\theta D,
	\end{equation}
while combining this angular size with a measurement of bolometric flux ($\fbol$) yields a direct value of the effective temperature \citep[see e.g.\@][]{code1976,boyajian2009,white2013}
	\begin{equation}
	\teff = \left( \frac{4 \fbol}{\sigma_{\mathrm{SB}}^{} \angdia^2} \right)^{1/4},
    \label{eq:teff}
	\end{equation}
where $\sigma_{\mathrm{SB}}$ is the Stefan-Boltzmann constant.

Interferometry is a powerful technique for stellar astrophysics as it only depends on stellar models to a small extent. However, due to seeing effects, optical transmission of the mirrors used, and the background photon noise, only the brightest stars on the sky can be observed using the interferometers today \citep{monnier2003}. 

\mystar (HD~181096, KIC~10005473) is an F6 subgiant star which has not been the subject of detailed studies before. It is a bright star with $V = 6.00$, meaning that it is bright enough for its angular size to be resolved using long-baseline optical interferometry.
\mystar was also within the field of view of the NASA \emph{Kepler} mission \citep{gilliland2010,koch2010} and as it exhibits solar-like oscillations, it can be studied using asteroseismology. 
Overall, these properties make \mystar an excellent test for the asteroseismic scaling relations, and it can help pave the way for a detailed analysis of a greater sample of subgiant stars.

The paper is organised as follows. In Sec.~\ref{sec:analysis} we provide an overview of the different observational methods and of how the different observed quantities are related to each other. We then proceed to go into more detail with each method. In Sec.~\ref{sec:results} we compute different stellar properties from the observations and we compare the different results obtained from the different techniques.
In Sec.~\ref{sec:models} we describe our stellar modelling efforts.
In Sec.~\ref{sec:discussion} we compare the observations with the results from stellar modelling and we demonstrate that the modelling seems to systematically underpredict the stellar radius. Finally we summarise our findings in Sec.~\ref{sec:conclusions}.

\section{Data analysis}
\label{sec:analysis}
\subsection{Overview of observables}
\label{ssub:overview}

\begin{figure}
	\centering
	\includegraphics{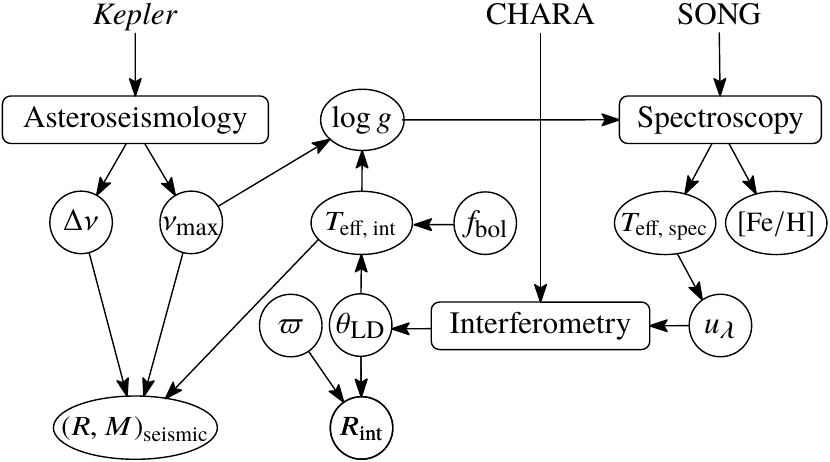}
	\caption{Flow diagram showing the relationships between the methods used and the derived stellar parameters.}
	\label{fig:flow}
\end{figure}

We perform an in-depth analysis of \mystar using interferometry (Sec.~\ref{sec:interferometry}), asteroseismology (Sec.~\ref{sec:asteroseismology}), spectroscopy (Sec.~\ref{sec:spectroscopy}), and grid-based stellar modelling (Sec.~\ref{sec:models}).  A graphic overview of the relationships between the variables and observational methods can be seen in Fig.\@~\ref{fig:flow}. Starting in the right-hand side of Fig.\@~\ref{fig:flow}, a literature value for the effective temperature determined from spectroscopy $T_{\textup{eff, spec}}$ is used to compute the first iteration of the linear limb-darkening coefficient $\ldc$. Combining the limb-darkening coefficient and the interferometric data, the limb-darkened angular diameter $\angdia$ of \mystar is found. Using a measured parallax $\varpi$ a distance can be derived and thus the linear radius of the star $\starradius_{\textup{int}}$ is then determined from Eq.~\ref{eq:linearradius}. Finally, from $\angdia$ and the bolometric flux of the star $\fbol$, an estimate of the effective temperature $T_{\textup{eff, int}}$ can be determined from interferometry using Eq.~\ref{eq:teff}.

We wish to compare the interferometric radius of the star with that predicted from asteroseismic inference. Using photometric data from \emph{Kepler}, the large frequency separation \dnu and the frequency of maximum power \numax can be computed. The logarithmic surface gravity $\log g$ can be estimated from Eq.~\ref{eq:numaxscalingrelation} using \numax and $T_{\textup{eff, int}}$. By anchoring $\log g$ in the spectroscopic analysis to this value, the metallicity \feh and a spectroscopic estimate of the effective temperature $T_{\textup{eff, spec}}$ can be determined. Then $T_{\textup{eff, spec}}$ can be fed back into a recalculation of the limb-darkening coefficient and the interferometric limb-darkened angular diameter \angdia. This calculation loop continues until no change in limb-darkening coefficient is found and consequently the calculated angular diameter remains unchanged from the last iteration.
An asteroseismic radius $\starradius_{\textup{seismic}}$ and an asteroseismic mass $\starmass_{\textup{seismic}}$ can be determined by combining asteroseismic parameters \dnu and \numax with an estimate of temperature ( Eqs.~\ref{eq:mscalingrelation}~\&~\ref{eq:rscalingrelation}).
Finally, we compare the measured physical parameters to the quantities from stellar modelling.

\subsection{Interferometry}
\label{sec:interferometry}
	\begin{table}
	\caption{Overview of PAVO interferometric observations.}
	\label{tab:obs}
	\begin{tabular}{lccc}
		\hline
		UT date & Calibrator$^1$ & Baseline$^2$  & No.\@ of scans\\ 
		\hline
		2013 July 8 	& acde 	& E2W1 & 4 \\
		2013 July 9 	& ace 	& S1W2 & 4 \\
		2014 Apr 8	 	& bf 	& E1W2 & 1 \\
		2014 Aug 16 	& cf 	& S2E2 & 5 \\
		2014 Aug 17 	& cf 	& E2W1 & 2 \\
		2014 Aug 18 	& cf$^3$& E2W2 & 3 \\
		\hline
		\multicolumn{4}{l}{\footnotesize{$^1$See Table~\ref{tab:calibrators}.}}\\
		\multicolumn{4}{l}{\footnotesize{$^2$The baselines have the following lengths:}}\\
		\multicolumn{4}{l}{\footnotesize{$\phantom{^2}$E2W2:~$156.27$~\si{\metre}; S1W2:~$210.97$~\si{\metre}; E1W2:~$221.82$~\si{\metre};}}\\ 
		\multicolumn{4}{l}{\footnotesize{$\phantom{^2}$S2E2:~$248.13$~\si{\metre}; E2W1:~$251.34$~\si{\metre}.}}\\
		\multicolumn{4}{l}{\footnotesize{$^3$The last scan was calibrated using only c.}}
	\end{tabular}
	\end{table}

	\begin{table}
	\caption{Calibrators used for the interferometric measurements. The uniform-disc angular diameter in the $R$-band is denoted {$\theta_{\mathrm{UD},R}$}}
	\label{tab:calibrators}
	\sisetup{round-mode=places, round-precision=3}
	\begin{tabular}{ll S[table-format=1.3] S[table-format=1.3] S[table-format=1.3] S[table-format=1.3] l}
		\hline
	HD & Sp. Type & {$V$} & {$K$} & {$E(B-V)$} & {$\theta_{\mathrm{UD},R}$} & ID \\
		\hline
	176131				& A2 V		& 7.08200 	& 6.74800 & 0.0068 & 0.154 & a\\ 
	176626				& A2 V		& 6.85200 	& 6.77100 & 0.0219 & 0.147 & b\\ 
	177003				& B2.5 IV	& 5.37700 	& 5.89500 & 0.0145 & 0.204 & c\\
    179095				& B8 IV 	& 6.91500	& 6.99000 & 0.0176 & 0.130 & d\\
	183142				& B8 V 		& 7.06900 	& 7.53400 & 0.0272 & 0.096 & e\\ 
	185872				& B9 III 	& 5.39900 	& 5.48000 & 0.0252 & 0.266 & f\\ 
		\hline
	\end{tabular}
	\end{table}
We measured the angular diameter of \mystar using long-baseline optical interferometry. We used the PAVO beam combiner \citep[Precision Astronomical Visible Observations;][]{ireland2008} at the CHARA array located at Mount Wilson Observatory, California \citep[Center for High Angular Resolution Astronomy;][]{ten2005}. 
The CHARA array consists of six 1-m telescopes in a Y-configuration, allowing 15 different baseline configurations between {34.07} and \SI{330.66}{\metre}. PAVO is a three-beam pupil-plane beam combiner, optimised for high sensitivity at visible wavelengths (${\sim}600$--$\SI{900}{\nano\metre}$).

Our observations were made using PAVO in two-telescope mode and baselines ranging from {157.27}--{251.34}~\si{\metre}. A summary of our observations can be found in Table~\ref{tab:obs}. Table~\ref{tab:calibrators} lists the six stars we used to calibrate the fringe visibilities of \mystar. Ideally an interferometric calibrator star is an unresolved point source with no close companions. The calibrator stars need to be observed as closely in time and in angular distance to the target object as possible in order to avoid changes in system variability, and therefore we observed the calibrator stars immediately before and after the target object. For all but one scan, the observing procedure was Calibrator~1 $\rightarrow$ Target $\rightarrow$ Calibrator~2. For the last scan of August 18 2014, only one calibrator was used as the second calibrator HD~185872 caused a miscalibration of target. This does not change the derived angular diameters.

The angular diameters of the calibrators were found using the $(V-K)$ surface brightness calibration of \citet{boyajian2014}. The $V$-band magnitudes were adopted from the Tycho-2 catalogue \citep{hoeg00}, and converted into the Johnson system using the calibration given by \citet{bessell2000}. $K$-band magnitudes were taken from the Two Micron All Sky Survey catalogue \citep{2mass}. Interstellar extinction was estimated from the dust map of \citet{green2015} and the extinction law of \citet{odonnell1994}. The calculated angular diameters were corrected for the limb-darkening to determine the corresponding uniform-disc diameter in $R$-band.

The data were reduced, calibrated, and analysed using the PAVO reduction pipeline, \citep[see e.g.\@][]{ireland2008,bazot2011,derekas2011,huber2012,maestro2013}. The uncertainties were estimated by performing Monte Carlo simulations with 100000 iterations assuming Gaussian uncertainties in the visibility measurements, $\SI{5}{\nano\metre}$ in the wavelength calibration, and $\SI{5}{\percentage}$ in the sizes of the calibrator stars.

We fitted a linear limb-darkened disc model to the visibility measurements $V_{\text{LD}}$ \citep{hanbury1974},
	\begin{equation}
	\label{eq:visibilityld}
	V_{\text{LD}} = \left( \frac{1 - \ldc}{2} + \frac{\ldc}{3} \right)^{-1}  \left(\frac{J_1(x)}{x} (1-\ldc) + \sqrt{\frac{\pi}{2}} \frac{J_{3/2}(x)}{x^{3/2}} \ldc \right),
	\end{equation}
where $x=\pi B \angdia\lambda^{-1}$. Here $\ldc$ is the linear limb-darkening coefficient, $J_n(x)$ is the $n$'th order Bessel function of the first kind, $B$ is the projected baseline, $\angdia$ is the limb-darkening corrected angular diameter, and $\lambda$ is the wavelength at which the observation was made. The product $B\lambda^{-1}$ is known as the spatial frequency.  

	\begin{figure}
\includegraphics[width=\columnwidth]{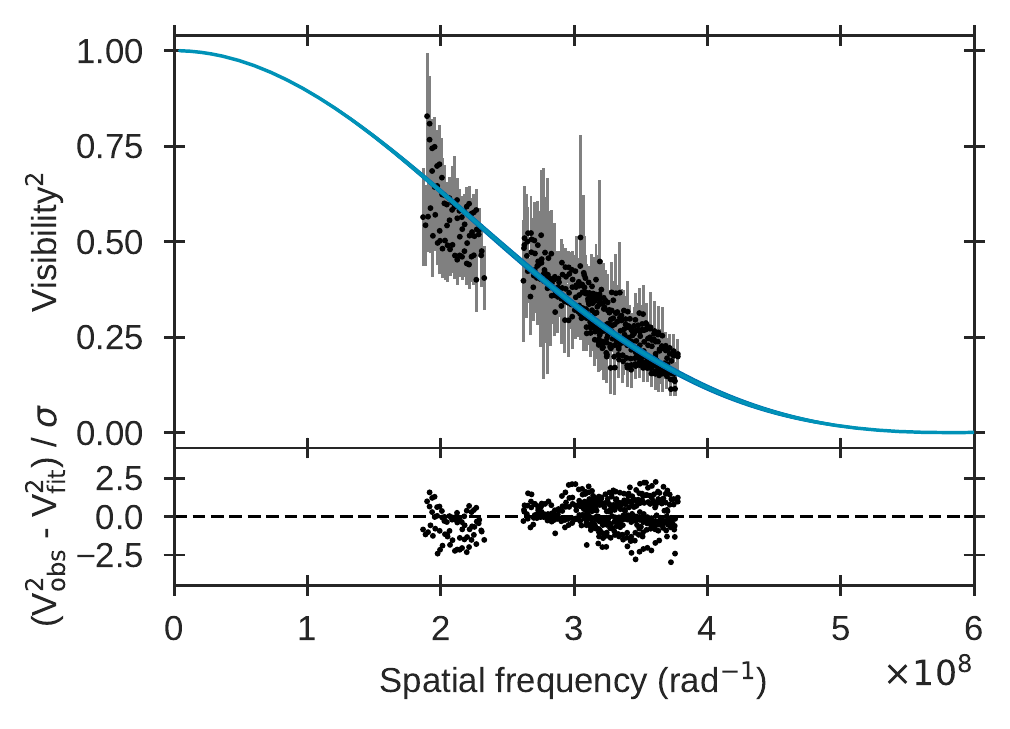}
	\caption{Interferometric measurements of \mystar from PAVO. The black dots with grey error bars show the squared fringe visibility measurements, while the blue curve shows the best-fitting limb-darkened disc model. The residuals weighted by the visibility uncertainties are shown in the bottom plot.}
	\label{fig:visibility}	
	\end{figure}

The limb-darkening coefficient \ldc of \mystar was estimated using a \teff-\ldc relation in the $R$-band (White et al., in prep.). The limb-darkening coefficient also has a metallicity and surface gravity dependence, but no strong relations with these quantities at these wavelengths were found and therefore our estimate of limb-darkening coefficient was found using only effective temperature.
The relation was found by performing 10000 iterations of a Monte Carlo simulation of the measured limb-darkening coefficients and temperatures from PAVO of 16 stars by allowing the values to vary within their uncertainties. The Sun was also added to the determination of the relation by using the limb-darkening coefficient from \citet{neckel1994}. Using the spectroscopic temperature (see Table~\ref{tab:par}), the limb-darkening coefficient for \mystar was determined to be $\ldc=\SI{0.22\pm0.05}{}$. Using this \ldc, the fit in Eq.\@~\ref{eq:visibilityld} to the visibility measurements yields a limb-darkened angular diameter of \mystar of $\angdia=\SI{0.443 \pm 0.007}{mas}$ (see Fig.\@~\ref{fig:visibility}).
When a uniform disc model, i.e.\@ a model that does not include limb darkening, is fitted to the data, then the uniform-disc angular diameter is found to be $\theta_{\mathrm{UD}}=\SI{0.435 \pm 0.005}{mas}$.

An interferometric measure of effective temperature $T_{\textup{eff, int}}$ can be found using an estimate of the bolometric flux at Earth.
The bolometric flux of \mystar was measured by \citet{casagrande2011} to be $\fbol = {(1.06 \pm 0.05)\times 10^{-7}}\si{ \erg \per\second \per\centi\meter\squared}$,
resulting in a effective temperature of $T_{\textup{eff, int}} = \SI{6350 \pm 90}{\kelvin}$. 

\subsection{Asteroseismology} 
\label{sec:asteroseismology}
The photometric time series of \mystar is available from the NASA \emph{Kepler} mission, which observed \mystar in short-cadence mode (${\sim}$~$\SI{1}{\minute}$) during quarter 15 (Q15) spanning $\SI{96.7}{\day}$ as part of \emph{Kepler} Guest Investigator Program GO40009. One safe mode event occurred during Q15, causing a gap in the photometric time series. Light curves were constructed from pixel data downloaded from the KASOC database\footnote{\url{www.kasoc.phys.au.dk}}. The raw time series was corrected for instrumental signals using the KASOC filter, which employs two median filters of different widths, with the final filter being a weighted sum of the two filters based on the variability in the light curve \citep{handberg2014}.
	\begin{figure}
	\includegraphics[width=\columnwidth]{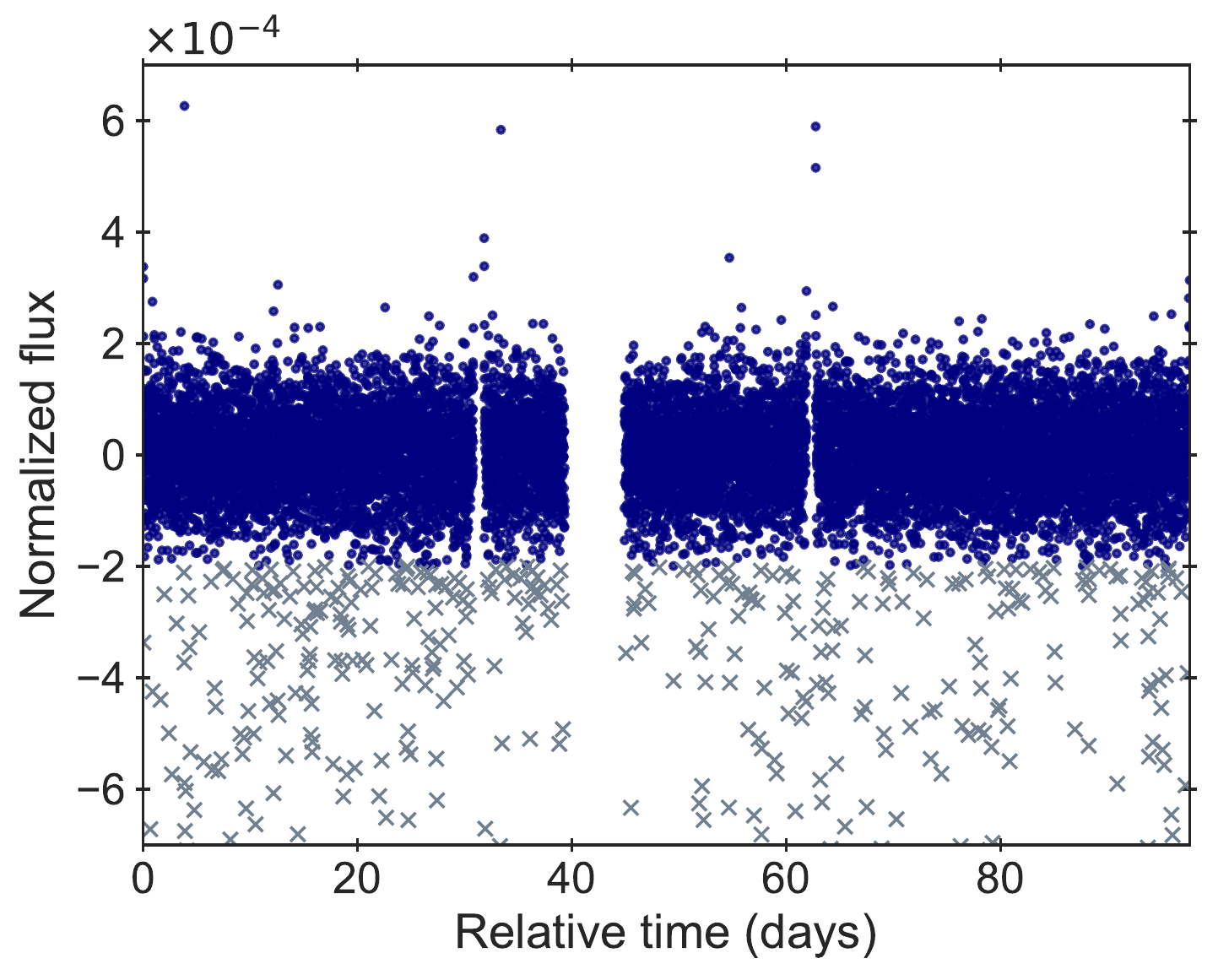}
	\caption{Q15 short-cadence time series of \mystar from \emph{Kepler} shown as blue points. Grey crosses show points ascribed to pointing jitter. For clarity, only \SI{10}{\percent} of the data are shown.}
	\label{fig:timeseries}
	\end{figure}
\begin{figure*}
	\includegraphics[width=\textwidth]{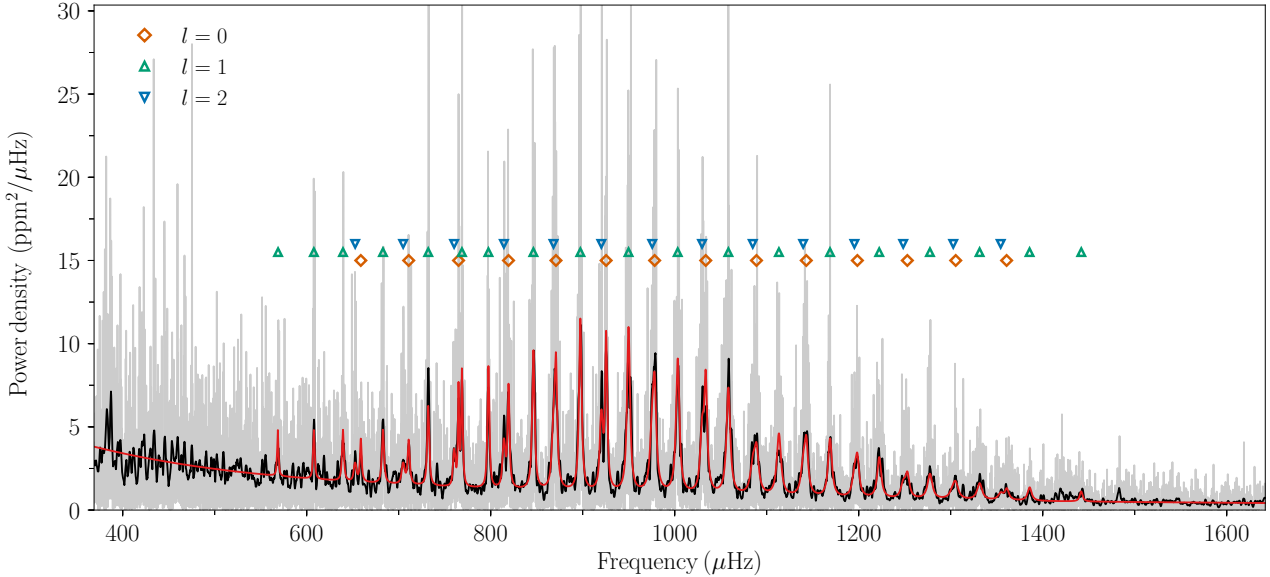}
	\caption{Power density spectrum of HR 7322. The full spectrum is shown in grey with a $\SI{3}{\micro\hertz}$ Epanechnikov smoothed version overlain in black. The fitted spectrum from the peak-bagging procedure is overlain in red. The markers indicate the frequency and angular degree of the fitted modes.}
	\label{fig:power}
	\end{figure*}

As seen in Fig.\@~\ref{fig:timeseries}, the time series of \mystar shows a substantial number of outliers below the average flux level. As the data were obtained a few months before the second reaction wheel of the spacecraft failed, we follow the same approach as \citet{johnson2014} and ascribe these outliers to pointing jitter caused by the increased friction that eventually led to the reaction wheel failure.
The power density spectrum (PDS; Fig.\@~\ref{fig:power}) used for further seismic analysis was constructed from a weighted least-squares sine-wave fitting, single-side calibrated, normalised according to Parseval's theorem, and converted to power density by multiplying by the effective observing length obtained from the integral of the spectral window \citep{kjeldsen1992}.

The individual mode frequencies for \mystar were extracted from the power spectrum using the peak-bagging approach described in \citet{lund2017}. Fig.\@~\ref{fig:power} shows the PDS with the frequency of the fitted modes indicated, and as seen here \mystar shows a departure from the regularity in the mode degree pattern around $\SI{780}{\micro\hertz}$ with two dipole modes (green triangles) being between two radial modes (orange diamonds) instead of only a single dipole mode.
First-guesses for the mode frequencies included in the peak-bagging were obtained from visual inspection of the PDS. We note that $l=1$ modes were treated in the same way as pure p-modes, but with amplitudes and linewidths decoupled from the $l=0$ modes.

The large frequency separation $\dnu$ and the frequency of maximum power $\numax$ were estimated by running the cleaned time series through the automated analysis pipeline described in \citet{huber2009,huber2011}, and they were determined to be $\dnu = \SI{53.92 \pm 0.20}{\micro\hertz}$ and $\numax = \SI{960 \pm 15}{\micro\hertz}$. The value of \numax is in agreement with a simple Lorentz fit to the amplitudes of the individual modes.

\citet{takeda2005} measured the projected rotational velocity $V \sin i$ of \mystar using spectroscopy to be \SI{3}{\kilo\meter\per\second}, indicating either a low rotation rate or a pole-on view. From the peak-bagging, no clear independent values can be obtained for the rotational splitting ($\nu_s$) or stellar inclination. However, a seismic equivalent for the projected rotational velocity can be derived from the projected rotational splitting $\nu_s \sin i$ and the modelled stellar radius as $V \sin i = 2\pi R \nu_s \sin i$ \citep{lund2014}. We find a value of $V \sin i = 4.5\pm 1.8\, \rm km\, s^{-1}$, in agreement with the value from \citet{takeda2005}.

\subsection{Spectroscopy}
\label{sec:spectroscopy}
The Hertzsprung SONG 1-m telescope \citep{andersen2014,grundahl2017} at Observatorio del Teide on Tenerife was used to obtain  high-resolution ($R = 90000$) \'echelle spectra of \mystar on March 13 and September 16, 2016. Extraction of spectra, flat fielding and wavelength calibration were carried out with the SONG data reduction pipeline. Individual spectra were combined in \textsc{iraf} after correction for Doppler shifts resulting in a spectrum in the ${\sim}4400$--$\SI{6900}{\angstrom}$ region with a signal-to-noise ratio of $\textup{S}/\textup{N}{\sim}400$ at $\SI{6000}{\angstrom}$. For this spectrum, equivalent widths of the spectral lines listed in \citet[Table~2]{nissen2015} were measured by Gaussian fitting to the line profiles. 

The equivalent widths were analysed with \textsc{marcs} model atmospheres \citep{gustafsson2008} with the method described in \citet{nissen2017} to obtain abundances of elements. 
As seen in Eq.\@~\ref{eq:numaxscalingrelation}, the frequency of maximum power is related to the surface gravity and the effective temperature $\numax \propto g / \sqrt{\teff}$. 
A logarithmic surface gravity of $\log g = 3.95 \pm 0.01$ was determined for \mystar by using the asteroseismic $\numax$ and the interferometric $\teff$ (see Table~\ref{tab:par})
and by adopting $\text{T}_{\text{eff},\odot} = \SI{5772}{\kelvin}$, $\log g_{\odot} = 4.438$, and $\nu_{\text{max},\odot} = \SI{3090}{\micro\hertz}$ for the Sun. Then, the spectroscopic $\teff$ was determined  from the requirement that the same Fe abundance should be obtained from \ion{Fe}{i} and \ion{Fe}{ii} lines. In this connection, non-LTE corrections from \citet{lind2012} were taken into account, which decreases $\teff$ by $\SI{50}{\kelvin}$ relative to the LTE value.
The results are $\teff = \SI{6313 \pm 35}{\kelvin}$ and $\feh = -0.23 \pm 0.04$. We assume that the systematic uncertainties are of the same order of magnitude as the statistical uncertainties and add these uncertainties in quadrature to get a combined uncertainty of $\SI{50}{\kelvin}$ and $\SI{0.06}{dex}$ respectively.
Comparing this effective temperature from spectroscopy with the effective temperature from interferometry, we see that they have an excellent agreement within ${\sim}0.4\sigma$. Using this spectroscopic $\teff$ value in the scaling relation does not change $\log g$ significantly, i.e.\@ by only $\SI{0.003}{dex}$. As the two temperatures agree, we choose to use the interferometric temperature in the following analysis. 

In addition, the ratio between the abundance of alpha-capture elements (Mg, Si, Ca, and Ti) and Fe was determined to be $\alphafe = 0.06 \pm 0.03$ showing that \mystar belongs to the population of low-$\alpha$ (thin disk) stars.

\section{Results}
\label{sec:results}

\begin{table}
	\caption{Summary of the mass and radius obtained using different asteroseismic scaling relations and the interferometric temperature $T_{\textup{eff, int}}$.}
	\label{tab:scalingrelations}
	\sisetup{round-mode=places, round-precision=3}
	\begin{tabular}{l S[table-format=1.2(2)] S[table-format=1.2(2)]}
		\hline
        							&	{$M$ [\si{\solarmass}]}	&	{\starradius [\si{\solarradius}]}	\\
        \hline
        Direct method   		& 	1.36 \pm 0.07			&	2.04 \pm 0.04 \\
        White et al. 2011		&	1.33 \pm 0.07			&	2.02 \pm 0.04 \\
        Sharma et al. 2016		&	1.27 \pm 0.07			&	1.98 \pm 0.04 \\
        Sahlholdt et al. 2018	&	1.25 \pm 0.06			&	1.98 \pm 0.04 \\
        Kallinger et al. 2018   &   1.27 \pm 0.07           &   2.00 \pm 0.04 \\
        Bellinger 2019           &   1.26 \pm 0.12          &   1.99 \pm 0.06 \\
        \hline
	\end{tabular}
\end{table}

\subsection{Interferometric radius}
 We know the limb-darkened angular diameter $\angdia$ of \mystar and thus the linear radius of the star can be estimated from Eq.~\ref{eq:linearradius} using an estimate of the distance, which is usually determined directly from the parallax $\varpi$. The parallax for \mystar was measured by the ESA missions \emph{Hipparcos} and \emph{Gaia} \citep{vanleeuwen2007,gaia2016,gaia2016dr1}. The \emph{Hipparcos} mission measured the parallax of \mystar to be $\varpi = \SI{23.79 \pm 0.32 }{mas}$, while the second data release for the \emph{Gaia} mission (\emph{Gaia} DR2) report the parallax of \mystar to be $\varpi = \SI{23.74 \pm 0.04 }{mas}$ \citep{gaiadr2}.
We assess the astrometric solution of \mystar from \emph{Gaia} by computing a re-normalised unit weight error of $\textup{RUWE}=0.82$ \citep{vo:gaia_dr2light}, which is surprisingly low for such a bright target owing primarily to the colour index $\textup{G}_{\textup{BP}}-\textup{G}_{\textup{RP}}=0.637$ and the number of good observations ($\textup{n}_{\textup{good}}=175$) for \mystar. Due to the good quality of the \emph{Gaia} five-parameter solution, we choose to rely on the \emph{Gaia} DR2 data for this bright target and trust a distance derived using this parallax.

We use the distance estimated for \mystar by \cite{bailerjones2018}, who inferred geometric distances from all parallaxes in the \emph{Gaia} DR2 catalogue by using a exponentially decreasing space density prior estimated from a model of the Milky Way. The model does not take stellar properties or reddening into account. A significant zero-point offset in \emph{Gaia} DR2 have been confirmed using various sources \citep[e.g.\@][]{arenou2018,zinn2018} in the sense that parallaxes from \emph{Gaia} DR2 are too small. \citet{bailerjones2018} take this account by incorporating the global parallax zero-point from \citet{lindegren2018} into the posterior probability density function.
Using the distance from \cite{bailerjones2018} in Eq.~\ref{eq:linearradius} yields an interferometric radius of the star of $\starradius = \SI{2.00 \pm 0.03}{\solarradius}$. 
An estimate of reddening of \mystar was found to be approximately zero using the three-dimensional dust map from \citet{green2018} and it does thus not violate the assumption of no reddening in the distance computation. This is also consistent with \mystar being close to us at a distance of ${\sim}\SI{43}{pc}$ and within the Local Bubble.

\subsection{Radius and mass from scaling relations}
\label{ssec:mass}
Assuming that \mystar is homologous to the Sun, the asteroseismic scaling relations Eqs.~\ref{eq:numaxscalingrelation}~\&~\ref{eq:dnuscalingrelation} are valid, and we can find the asteroseismic estimate using the direct method in Eq.~\ref{eq:rscalingrelation}. We use the same $\nu_{\text{max},\odot}$ as in Sec.~\ref{sec:spectroscopy} along with $\dnu_{\textup{solar}}=\SI{135.1}{\micro\hertz}$. We find the asteroseismic radius and mass of \mystar to be $\starradius=\SI{2.04 \pm 0.04}{\solarradius}$ and $\starmass = \SI{1.36 \pm 0.07}{\solarmass}$ using the direct method.

However, the homology assumption leading to the scaling relation is not strictly valid. It has been shown that Eq.\@~\ref{eq:rscalingrelation} holds within $\SI{5}{\percentage}$ for dwarfs, subgiants, and giants \citep[see e.g.\@][]{stello2009,white2013,huber2017} and that Eq.~\ref{eq:mscalingrelation} holds within $10-\SI{15}{\percentage}$ \citep[see e.g.\@][]{miglio2012,chaplin2014}. 

Previous studies have found that the scaling relations for \dnu and \numax can be improved using corrections, typically written as a correction factor \fnumax and \fdnu multiplied on the right hand side of Eqs.~\ref{eq:numaxscalingrelation}~\&~\ref{eq:dnuscalingrelation} respectively.
We still do not have a complete physical understanding of \numax, which also means that determining \fnumax is an unresolved issue. \citet{belkacem2011} proposed a dependence on the Mach number and mixing length parameter in the near-surface layers, meaning that \numax should depend heavily on the physical conditions near the surface of the star.
\citet{viani2017} find that the most visible deviation between the observations and stellar models can be explained by adding a dependency on the mean molecular weight, however it seems that \feh of \mystar is within the metallicity range where this additional term has little-to-no influence.
\citet{white2011} studied a grid of stellar models and suggested a second-order polynomial correction to the \dnu scaling relation (Eq.\@~\ref{eq:dnuscalingrelation}) where \fdnu is a function of the effective temperature. \citet{sharma2016} suggested a correction depending on the evolutionary state, \feh, $\log g$, and \teff, where \fdnu is found by interpolation in a grid based on stellar modelling. \citet{sahlholdt2018} studied about $100$ main-sequence stars from the {\it Kepler} LEGACY sample \citep{lund2017,silvaaguirre2017} and the Kages sample \citep{davies2016,silvaaguirre2015} and found a linear and a quadratic polynomial parametrisation of \fdnu and \fnumax respectively, both only depending on the effective temperature.

A different approach is to derive new scaling relations with different exponents than in Eqs.~\ref{eq:mscalingrelation}~\&~\ref{eq:rscalingrelation}.
\citet{bellinger2019} derive new scaling relations not only for radius and mass but for stellar age as well, based on the same sample of main-sequence stars as \citet{sahlholdt2018}. In contrast to \citet{sahlholdt2018}, these scaling relations for mass and radius depend not only on effective temperature, but also contain a small, but non-zero dependency on metallicity as well.
\citet{kallinger2018} derive new non-linear scaling relation based on six red giants in eclipsing binary system along with about 60 red giants in the two open clusters NGC~6791 and NGC~6819 and as their correction is purely empirical, they do not contain any model-based correction terms.

In Table~\ref{tab:scalingrelations}, the radius and mass of \mystar are calculated using these corrections, and we see that the different versions of the asteroseismic scaling relation for radius have a maximum ${\sim}0.8\sigma$ difference from the interferometric radius and are thus all in good agreement with the radius from interferometry. 

\subsection{Stellar modelling}
\label{sec:models}
\begin{figure*}
	\includegraphics[width=\textwidth]{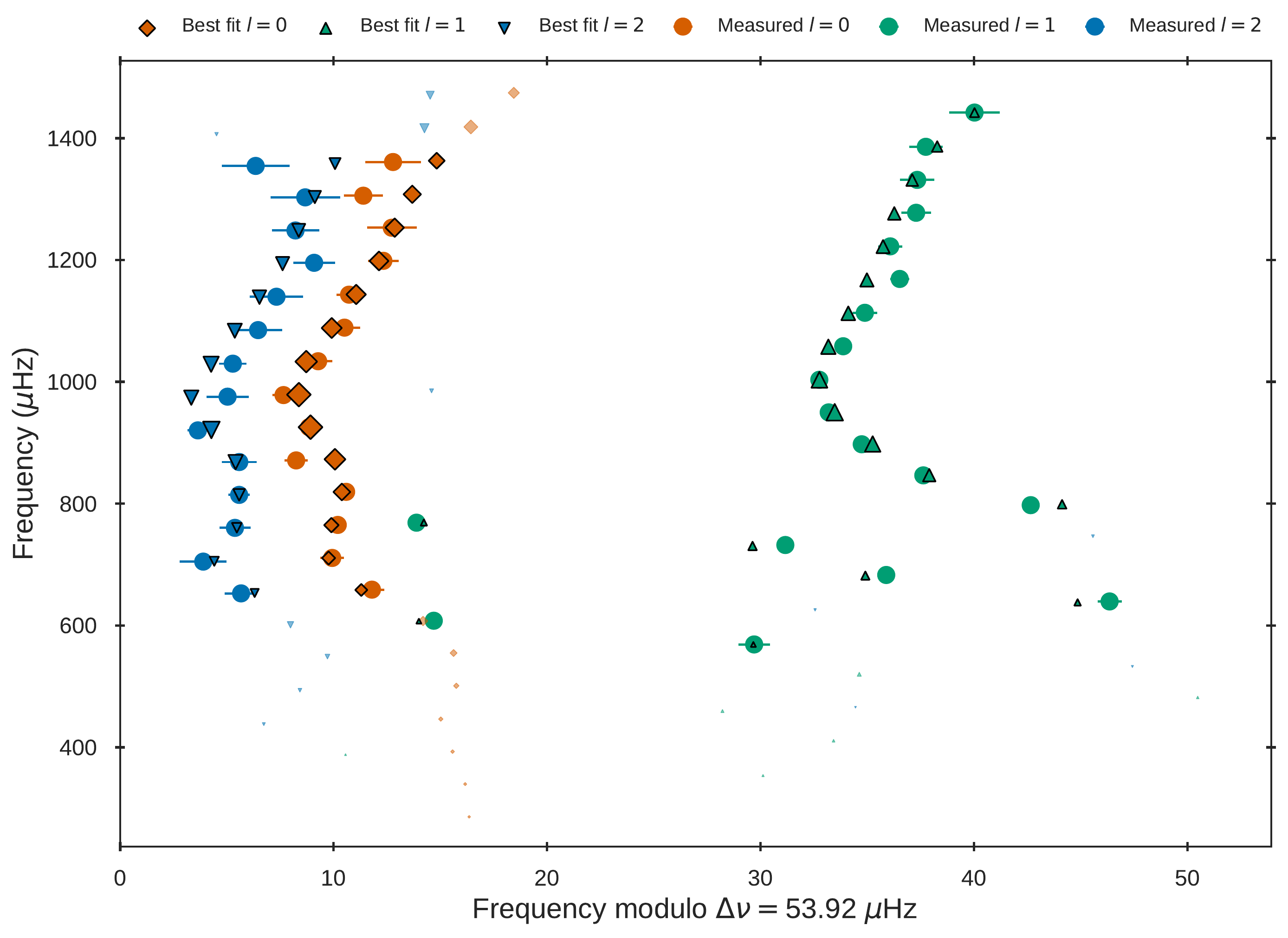}
	\caption{\'Echelle diagram of \mystar. The coloured circles show the observed oscillation modes (red: $l=0$, green $l=1$, blue: $l=2$), while the coloured symbols with a black outline show the modes predicted from the best-fitting model from \emph{lowmlt} (same colour coding), corrected using the surface correction from \citet{ball2014}. The size of the symbols from the model are scaled inversely with their normalised mode inertias which is correlated to the mode amplitudes and linewidths \citep{benomar2014}: the larger the symbol, the greater the probability of the mode being observed. The lighter coloured symbols with no outline are not matched to any observation, but are still predicted by the model.}
	\label{fig:echelle}
\end{figure*}

\begin{figure*}
	\includegraphics[width=\textwidth]{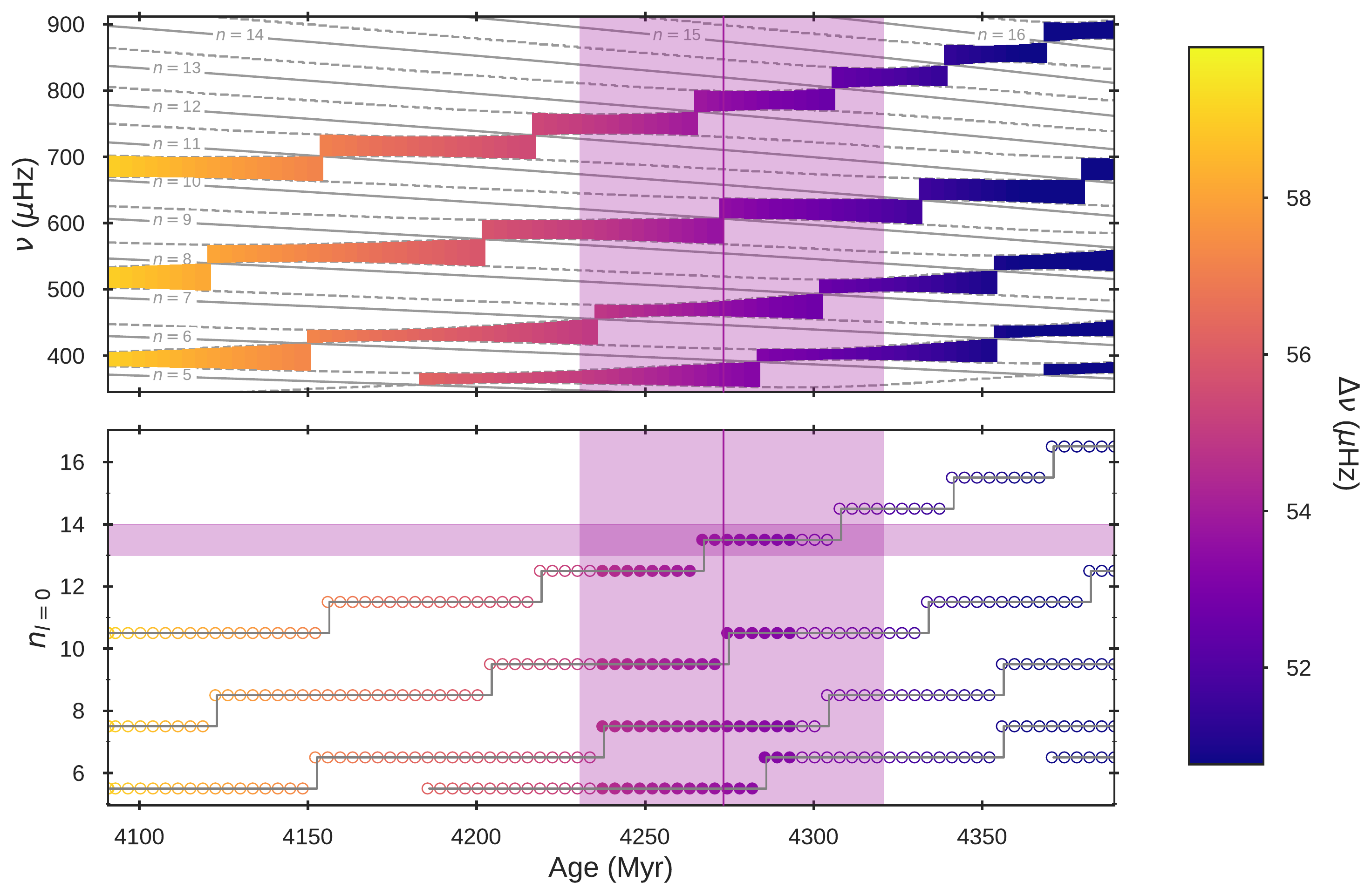}
	\caption{Frequency pattern changes as a function of stellar ages. Models are from a track in the refined \emph{lowmlt} grid with a given stellar mass and metallicity. Only every tenth model is plotted for clarity.
    Upper panel: The evolution of the frequencies of radial modes ($l=0$, solid lines) and the dipole modes ($l=1$, dashed lines) as a function of stellar age. The coloured area in-between two dipole lines at a given age marks where two dipole modes fall in-between two radial modes and thus marks the irregularities in the otherwise evenly-spaced pattern. The colour corresponds to the \dnu at the given age. The vertical magenta line marks the age of the best-fitting model in this grid and the vertical magenta-shaded areas the quoted uncertainties on stellar age from the fit (see Table~\ref{tab:par}). 
    Lower: Same models as in the upper panel, but now the second axis shows the radial order $n$ of the $l=0$ modes which have 2 $l=1$ modes in-between them. The filled circles indicate models where $\Delta \nu$ is within 5$\sigma$ of the observed. The empty circles show the remainder of the models. The vertical red line and red-shaded area show the same as for the upper panel, while the horizontal red shaded area shows the position of the observed avoided crossing between $l=0$ $n=13$ and $n=14$.}
	\label{fig:mmlifetimes}
\end{figure*}
We determine stellar properties of \mystar using grid-based stellar modelling. We construct grids of theoretical models of stellar evolution covering the necessary parameter space, which we compare the observed parameters to the predicted theoretical quantities. Quantities like stellar age can thus be estimated from the fit to the other observables. In this case, we fit the observed individual frequencies of \mystar along with the spectroscopic metallicity and the interferometric temperature to the grid.

Grids of stellar models were computed with the Garching Stellar Evolution Code \citep[\garstec;][]{weiss2008}. The stellar models are computed using the \citet{grevesse1998} solar mixture along with the OPAL equation of state \citep{rogers2002} and OPAL opacities \citep{opal96} at high temperatures supplemented by the opacities of \citet{ferguson2005} at low temperatures.
\garstec uses the NACRE nuclear reaction rates \citep{angulo1999} except for $^{14}$N($p,\gamma$)$^{15}$O and $^{12}$C($\alpha,\gamma$)$^{16}$O for which the rates from \citet{formicola2004} and \citet{hammer2005} were used.
Convection was treated using the mixing-length formalism \citep{kippehahn2012}, where the fixed mixing length parameter was set to a solar-calibrated value of $\amlt=1.818$.
In the modelling of \mystar, we used an Eddington grey atmosphere. Diffusion and settling of helium and heavier elements were not included, neither was convective overshooting. The stellar grid samples masses from $1.0$--$1.5$~\si{\solarmass} in steps of \SI{0.01}{\solarmass} and it samples metallicities from $\meh=-0.32$ to $-0.14$ in steps of $0.03$, assuming a fixed linear Galactic chemical evolution model of $\Delta Y / \Delta Z = 1.4$ \citep{balser2006}. The grid covers \dnu in the range $50$--$60$~\si{\micro\hertz}, thus spanning the parameter space from about $\SI{5}{\micro\hertz}$ on both sides of the observed \dnu. The frequencies of the stellar models were computed using the Aarhus adiabatic pulsation code \citep[\adipls;][]{christensen2008}. In order to correct for the systematic difference between the observed and the calculated frequencies introduced by the erroneous treatment of the near-surface layer in current stellar models, the two-term surface correction from \citet{ball2014} was applied to the computed frequencies.

Three additional grids of stellar models were made in the same manner as described above (which we nickname \emph{nor}): one with a lower mixing length parameter of $\amlt = 1.6$ (\emph{lowmlt}), one with a higher mixing length parameter of $\amlt = 2.0$ (\emph{highmlt}), and one including exponential convective overshooting with an efficiency parameter of $f = 0.016$ (\emph{ove}) \citep{freytag1996, weiss2008}.

We use the BAyesian STellar Algorithm \citep[\basta;][]{silvaaguirre2015,silvaaguirre2017} to determine the stellar parameters of \mystar. Given a precomputed grid of stellar models, \basta uses a Bayesian approach to compute the probability density function of a given stellar parameter using a set of observational constraints. 
\basta allows the possibility to add prior knowledge to the Bayesian fit, and we used the Salpeter Initial Mass Function \citep{salpeter1955} as a prior to quantify our expectation of most stars being low-mass stars.

Including the frequencies in the fit has been done for other subgiant stars such as $\eta$~Boo \citep{kjeldsen1995b,christensen1995, carrier2005}, $\beta$~Hyi \citep{bedding2007,brandao2011}, and more recently $\mu$~Her \citep{grundahl2017, li2019} and TOI~197 \citep{huber2019}. As mentioned in Sec.~\ref{sec:introduction}, the observed p-modes in main-sequence stars are approximately evenly spaced in frequency. However, post-main-sequence stars do show deviations from this regularity as mode bumping occur due to a coupling between the acoustic p-modes in the expanding convective envelope and the buoyancy-driven g-modes in the stellar core \citep{osaki1975,aizenman1977}. This coupling causes a range of non-radial pulsation modes to get mixed character -- behaving like p-modes in the envelope and like g-modes in the core -- 
which shifts the mode frequencies from their regular spacing and makes them stand out in the \'echelle diagrams as an avoided crossing.

In the \'echelle diagram of \mystar (Fig.~\ref{fig:echelle}), one dipole ($l=1$) avoided crossing is clearly visible.
Fig.~\ref{fig:mmlifetimes} shows how powerful avoided crossings can be to estimate precise stellar age. For a given mass and metallicity, only a few models spanning a narrow range in age of approximately $30$~Myr have the avoided crossing in-between the radial modes of radial order $n=13$ and $n=14$. When this is expanded to include the other masses and metallicities of the grid, the age range spans only about $100$~Myr, underlining how fitting this signature affects the uncertainty in age by constraining the parameter space much more than the spectroscopic values or the global asteroseimic parameters. This allows precise determination of stellar parameters \citep[see e.g.][]{deheuvels2011,benomar2014} and makes it possible to measure the age of the star with a relative statistical uncertainty of a few per~cent \citep{metcalfe2010,tian2015}. 

\begin{table}
	\caption{Summary of the results from the different grids of stellar models.}
	\label{tab:modelresults}
	\sisetup{round-mode=places, round-precision=3}
	\begin{tabular}{l S[table-format=1.3] S[table-format=1.3] S[table-format=1.3] S[table-format=1.3]}
		\hline
		{} 		        		&	{\centering\emph{nor}}				&	{\centering\emph{lowmlt}}			&   {\centering\emph{highmlt}}          &	{\centering\emph{ove}}	\\[3pt]
		\hline
		{\starmass (\si{\solarmass})}		& 	{$1.24\substack{+0.01 \\ -0.01}$}	&	{$1.19\substack{+0.01 \\ -0.01}$}	&   {$1.27\substack{+0.01 \\ -0.01}$}	&	{$1.25\substack{+0.01 \\ -0.01}$} \\[3pt]
		{\starradius (\si{\solarradius})}	& 	{$1.967\substack{+0.005 \\ -0.01}$}	&	{$1.94\substack{+0.008 \\ -0.008}$}	&   {$1.98\substack{+0.002 \\ -0.008}$}	&   {$1.967\substack{+0.009 \\ -0.01}$}	\\[3pt]
        {\teff (K)}  			        	& 	{$6440\substack{+35 \\ -5}$}		&	{$6295\substack{+40 \\ -21}$}		&   {$6600\substack{+45 \\ -2}$}	    &   {$6495\substack{+19 \\ -23}$}	\\[3pt]
        {\feh (dex)}			        	&	{$-0.20\substack{+0.03 \\ -0.03}$}	&	{$-0.23\substack{+0.03 \\ -0.03}$}	&   {$-0.23\substack{+0.03 \\ -0.03}$}	&   {$-0.20\substack{+0.03 \\ -0.03}$} \\[3pt]
        {Age (Myr)}					        &	{$3834\substack{+38 \\ -36}$}		&	{$4333\substack{+44 \\ -55}$}		&   {$3420\substack{+20 \\ -76}$}	    &   {$3611\substack{+81 \\ -43}$}	\\[3pt]
        \hline
	\end{tabular}
\end{table}

The results extracted from the probability density functions for each grid when fitting metallicity, interferometric temperature, and individual frequencies are seen in Table~\ref{tab:modelresults}. The metallicity of \mystar was computed using the spectroscopic \feh and \alphafe following \citet{salaris2002}.
We compared the frequencies of the model in each grid with the lowest $\chi^2$ value to the observed oscillation frequencies. All four best-fitting models make very reasonable fits to the individual frequencies, but the set of frequencies from the best-fitting model in the \emph{lowmlt} grid was the best match. An \'echelle diagram of the observed frequencies and the modelled frequencies from the best-fitting model from the \emph{lowmlt} grid is seen in Fig.\@~\ref{fig:echelle}.
We note that the model reproduces the clear dipole avoided crossing near \SI{780}{\micro\hertz} along with the two lower $l=1$ modes, which also seem to be mixed. 
There seems to be a frequency bump near \SI{1200}{\micro\hertz} that is present both in the $l=1$ and $l=2$ modes, which the modelled frequencies do not reproduce. As this bump appears in a part of the power spectrum with a high signal-to-noise ratio, this could be a real physical signature. This bump can be a result of the mismatch between the helium glitch signature in the observed and modelled frequency pattern \citep[see e.g.]{verma2017}. A mismatch would be of the same magnitude as the observed difference of $\sim{}$\SI{1}{\micro\hertz} and it would not depend on the spherical degree $l$. The helium glitch signature is difficult to measure for subgiants because of the avoided crossings, and therefore we have not added the helium glitch signature as a constraint to the fit.

The result from the \emph{lowmlt} grid best reproduces all the observational constraints. In particular, the effective temperature of the other grids disagree significantly with the observed effective temperatures, given in Table~\ref{tab:par}.
This is an effect of the different mixing length parameter \amlt in the grids. 
A decrease in \amlt changes the behaviour of the evolutionary tracks in a similar way to a decrease in metallicity by shifting the tracks towards hotter temperatures. This can explain why the \emph{lowmlt} grid with the sub-solar \amlt finds a solution with a temperature about $\SI{150}{\kelvin}$ lower than the \emph{nor} and \emph{ove} grids with a solar \amlt and about $\SI{300}{\kelvin}$ lower than the \emph{highmlt} grid with a super-solar \amlt.

The value of the mixing length parameter \amlt is known to vary across the H-R diagram \citep{trampedach2014, magic2015, mosumgaard2018}. 
The \stagger grid \citep{magic2015} predicts the mixing length parameter of \mystar based on the interferometric temperature, spectroscopic metallicity, and $\log g$ to be less than their solar-calibrated value by about $\Delta \amlt = 0.2$, in agreement with the \amlt in the favoured \emph{lowmlt} grid.

Observed individual frequencies constrain the parameter space considerably, narrowing down the formal uncertainties of the different parameters in the fit. In the case of mass and metallicity, we see in Table~\ref{tab:modelresults} that the uncertainties are equal to the resolution of the grid.
In order to check if the uncertainties were physical or due to the limited grid resolution, we recomputed a finer grid with a higher resolution in mass $\Delta M=0.001\si{\solarmass}$ and metallicity $\Delta \feh = 0.01$. The results from this higher resolution grid can be seen in Table~\ref{tab:par}, and a plot of the probability density functions of the refined \emph{lowmlt} grid can be seen in Fig.~\ref{fig:corner} in appendix.
Interestingly, the solution only changed marginally and the uncertainties are at the same order of magnitude as in the results from the coarser version of the grid, showing that these narrow statistical uncertainties are due to the constrained parameter space and not due to the resolution of the grids.

\begin{figure}
	\includegraphics[width=\columnwidth]{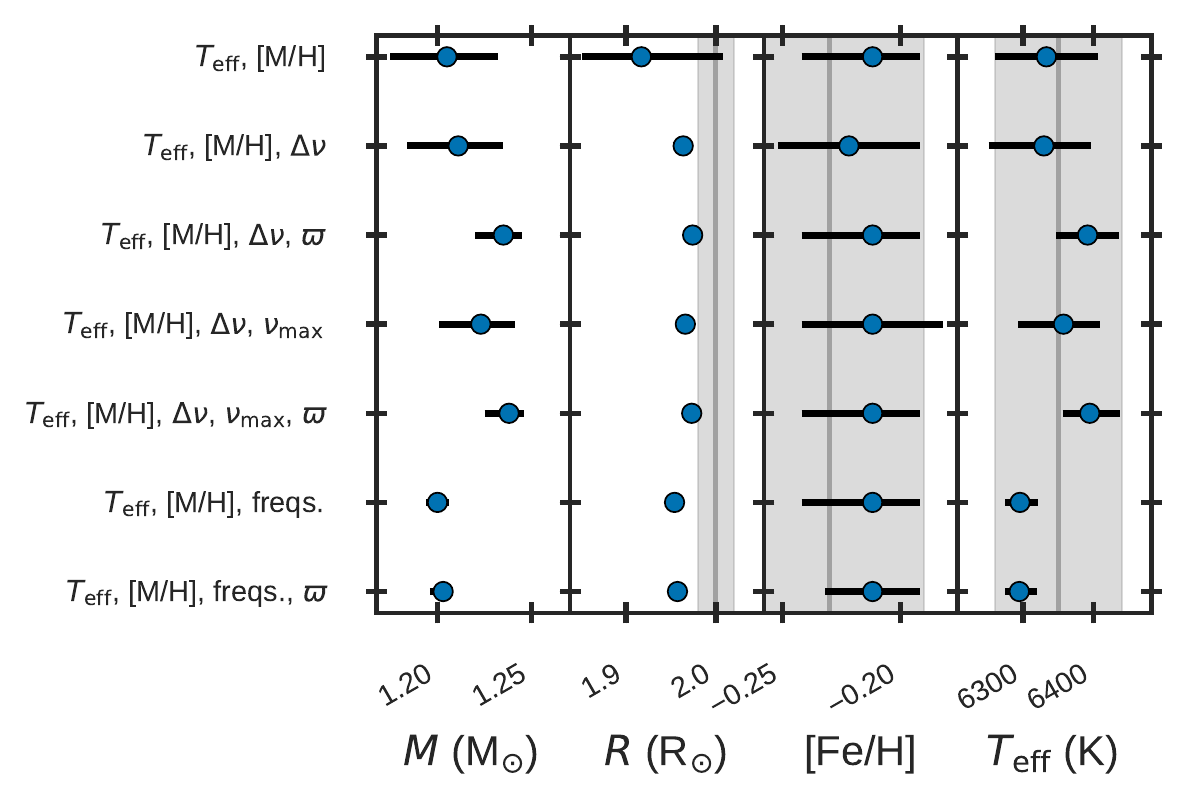}
	\caption{The median, 16th, and 84th quantile of the probability density functions obtained when fitting different sets of observables to the refined \emph{lowmlt} grid. The grey lines and areas mark the observed values. When we added the \emph{Gaia} parallax $\varpi$ as a constraint, we used the 2MASS H magnitude of \mystar \citep{2mass}.}
	\label{fig:fitstorefinedgrid}
\end{figure}

In Fig.~\ref{fig:fitstorefinedgrid}, we see the results of fitting different sets of parameters to this refined \emph{lowmlt} grid. We see that by supplementing spectroscopic parameters with asteroseismic ones, we narrow down the uncertainties in particular in radius. However, we also see that none of the fits containing asteroseismic constraints reproduces the interferometric radius within 1~$\sigma$.

\section{Discussion}
\label{sec:discussion}
\begin{figure}
	\includegraphics[width=\columnwidth]{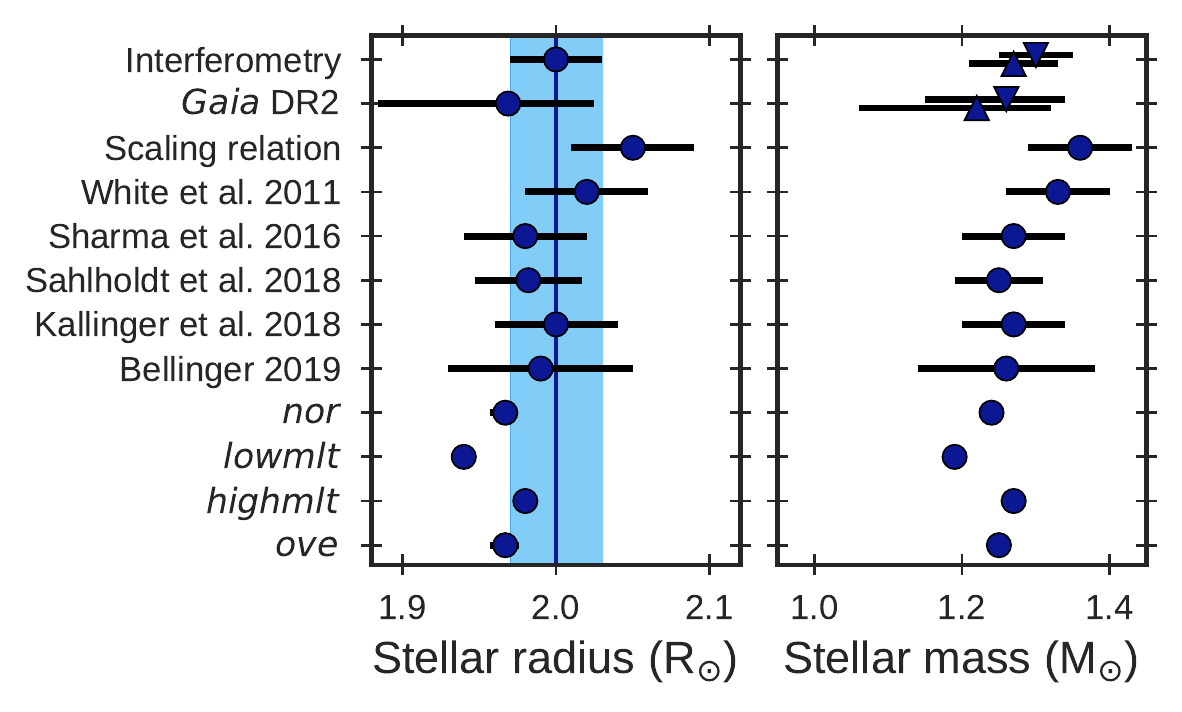}
	\caption{A visual comparison of the different stellar radius and mass estimates. The vertical blue line and band represent the interferometric radii and the $1\sigma$ uncertainties. The points with error bars show the different other estimates discussed in the text and in Tables~\ref{tab:scalingrelations}~and~\ref{tab:modelresults}. The mass estimates from interferometry and from \emph{Gaia}~DR2 are computed directly from Eqs.~\ref{eq:numaxscalingrelation} (triangles pointing down) and \ref{eq:dnuscalingrelation} (triangles pointing up) using the global asteroseismic parameters and the interferometric effective temperature.}
	\label{fig:comparison}
\end{figure}

If we compare the stellar radii obtained from the direct method and corrections to it (Table~\ref{tab:scalingrelations}) to the stellar radii obtained by modelling the individual frequencies (Table~\ref{tab:modelresults}), we get a maximum difference of ${\sim}2.5\sigma$, see also Fig.~\ref{fig:comparison}. The radii from fitting the frequency pattern to stellar models are all systematically lower than the radii derived from asteroseismic scaling relations and interferometry, and the percentage difference between the radius from the stellar modelling and the observations is in all but one case larger than what can be ascribed to statistical and systematic uncertainties from the chosen input physics \citep{silvaaguirre2015}.

We explored possible causes for the radii from the stellar modelling being systematically lower than the observed. 
As discussed in the previous section, we varied the mixing length parameter between the different grids. The tension between fitting the interferometric temperature and radius simultaneously is clear: The \emph{highmlt} grid could fit the radius within $0.7\sigma$, while the temperature was $2.8\sigma$ off, and the \emph{lowmlt} grid could fit the temperature within ${\sim}0.5\sigma$ but is off by ${\sim}1.7\sigma$ in radius. 

We decreased the fixed helium enrichment law used in the computations of all four grids of stellar models to $\Delta Y / \Delta Z = 1$, which caused the resulting effective temperature to decrease with about ${\sim}\SI{40}{\kelvin}$. The stellar radius did not change from the results in Table~\ref{tab:modelresults}.
We explored the effect of varying the initial helium abundance $Y_{\mathrm{0}}$ pseudorandomly within a grid and found that only by allowing $Y_{\mathrm{0}}$ to be less than the primordial helium abundance we could get a radius close to $\SI{2.00}{\solarradius}$ from the models. This is not the first time that helium abundances below the standard big bang nucleosynthesis value are favoured in asteroseismic analysis of solar-like oscillators \citep{metcalfe2010,mathur2012} and the reasons for this degeneracy has yet to be understood. 

We added the interferometric radius as an additional constrain during the fit, which slightly increases the mass and the radius by at most $\SI{0.01}{\solarmass}$ and $\SI{0.01}{\solarradius}$ respectively while decreasing the effective temperature by around $\SI{20}{\kelvin}$ for all four grids. That a discrepancy remains between the input radii and the obtained one is due to the fact that the individual frequencies are the main contributor to the likelihood computation.

\begin{table*}
    \centering
	\caption{Summary of the measured stellar parameters for \mystar along with the results of the grid-based stellar modelling.}
	\label{tab:par}
	\sisetup{round-mode=places, round-precision=3}
	\begin{tabular}{l S[table-format=1.3] S[table-format=1.3] S[table-format=1.3] S[table-format=1.3] S[table-format=1.3]}
		\hline
		{Stellar parameter}				& {Spectroscopy} 		& {Interferometry} 			&	{Asteroseismology}	&	{Direct method$^a$} 	&	{Modelling$^b$}	\\[3pt]
		\hline
        \starmass (\si{\solarmass})		& {\centering\dots}		& {\centering\dots}			& {\centering\dots}		&	\SI{1.35 \pm 0.07}{}	&	{$1.200\substack{+0.006 \\ -0.006}$}	\\[3pt]
		\starradius (\si{\solarradius})	& {\centering\dots}		& \SI{2.00 \pm 0.03}{}		& {\centering\dots}		&	\SI{2.04\pm0.04}{}		&	\SI{1.954 \pm 0.006}{}	\\[3pt]
        $L$ (\si{\solarlum})        	& {\centering\dots}		& {\centering\dots}			& {\centering\dots}		&	{\centering\dots}	 	&	\SI{5.37 \pm 0.06}{}	    \\[3pt]
		$\log g$ 						& {\centering\dots} 	& {\centering\dots}			& {\centering\dots}		&	\SI{3.95 \pm 0.01}{}	&	\SI{3.936 \pm 0.001}{}	\\[3pt]
		$\feh$ 							& \SI{-0.23 \pm 0.06}{} & {\centering\dots} 		& {\centering\dots}		&	{\centering\dots}		&	{$-0.21\substack{+0.02 \\ -0.03}$}    \\[3pt]
		$\teff$ (\si{\kelvin}) 			& \SI{6313 \pm 50}{}	& \SI{6350 \pm 90}{} 		& {\centering\dots}		&	{\centering\dots} 		&	{$6295\substack{+26 \\ -22}$}	\\[3pt]
 		Age (Myr)						& {\centering\dots}		& {\centering\dots}			& {\centering\dots}		&	{\centering\dots}		&	{$4273\substack{+47 \\ -43}$}	\\[3pt]
		\ldc 							& {\centering\dots}		& \SI{0.22\pm0.05}{}		& {\centering\dots}		&	{\centering\dots}		&	{\centering\dots}	\\[3pt]
		\angdia (\textup{mas}) 			& {\centering\dots}		& \SI{0.443 \pm 0.007}{}	& {\centering\dots}		&	{\centering\dots}		&	{\centering\dots}	\\[3pt]
		\numax (\si{\micro\hertz})		& {\centering\dots}		& {\centering\dots}			& \SI{960 \pm 15}{}	& 	{\centering\dots}		&	\SI{932 \pm 2}{}$^c$	\\[3pt]
		\dnu (\si{\micro\hertz})		& {\centering\dots}		& {\centering\dots}			& \SI{53.9 \pm 0.2}{}	& 	{\centering\dots}		&	\SI{53.8 \pm 0.1}{}	\\[3pt] 
		\hline
        \multicolumn{5}{l}{\footnotesize{$^a$Results using the interferometric temperature.}} \\
		\multicolumn{5}{l}{\footnotesize{$^b$Results from the refined \emph{lowmlt} grid (see Fig.\@~\ref{fig:echelle}).}} \\
		\multicolumn{5}{l}{\footnotesize{$^c$Computed from the scaling relation and the Stefan-Boltzmann law using $R$, $L$, and \teff in the models.}} \\
        \end{tabular}
	\end{table*}

As discussed in Sec.\@~\ref{sec:models}, the currently used one-dimensional stellar evolutionary codes do not treat the outermost layers of the star adequately, giving rise to the need of a surface correction in order to correct for the systematic differences between the observed and the modelled frequencies.
Advances in stellar modelling have made it possible to replace the outermost layers of the star with a patch of three-dimensional atmospheres \citep[e.g.\@][]{jorgensen2017}.
By changing the physics in the outer envelope of the star, the outer boundary of the star changes and the radius increases. However, even for a subgiant like \mystar, which does have a larger outer super-adiabatic layer than the main-sequence stars studied in \citet{jorgensen2017}, this effect only shifts the photosphere by $650$--$\SI{1400}{\kilo\metre}$ corresponding to at most $\SI{{\sim}0.002}{\solarradius}$ (J{\o}rgensen, private communication).

We find no model parameter that solves the systematic offset between the models and the interferometric observations.
The interferometric analysis contains a dependency on stellar models: how we estimate and model limb darkening. 
If the linear limb-darkening coefficient is overestimated, then angular diameter would be slightly smaller, making the interferometric radius smaller and the agreement between the two methods better. A smaller angular diameter would also increase the interferometric estimate of the effective temperature, improving the agreement with spectroscopy. 

Even though the radius from grid-based modelling is systematically lower than the observations, 
the radius obtained from the revised scaling relation adopting the correction from \citet{sharma2016}, \citet{sahlholdt2018}, \citet{kallinger2010} or \citet{bellinger2019} agree nicely with both the radius from interferometry, see Fig.~\ref{fig:comparison}. The corrections from \citet{sahlholdt2018} predict a mass and radius that is almost identical to the ones found in the best-fitting model in \emph{nor}. This is not surprising as \citet{sahlholdt2018} found their correction based on stellar parameters from \basta with a mixing length parameter \amlt very close to the one in \emph{nor} and \emph{ove}.

\emph{Gaia} DR2 also provides an estimate of radius by inferring the luminosity and temperature from wide-band photometry and their measure of parallax, using the Stefan-Boltzmann law to get a radius of \mystar of $\starradius=1.97\substack{+0.06 \\ -0.09}\si{\solarradius}$ \citep{andrae2018}.

If we compare the masses from scaling relations to the masses of the models in Table~\ref{tab:modelresults}, we again see that the latter predict lower values than those from scaling relations (see Fig.~\ref{fig:comparison}). We tried relaxing our prior assumption about the initial mass function, but it did not change our conclusions as the strongest constraints to the results from our modelling efforts come from the individual frequencies.

From Table~\ref{tab:scalingrelations}, it is clear that the difference in mass between the values computed from scaling relations listed is less than $\SI{10}{\percentage}$. The difference between the mass from modelling and the mass estimate from Eq.\@~\ref{eq:mscalingrelation} is about $\SI{13}{\percentage}$ or just above $2\sigma$, which is consistent with the offset also seen in radius between the modelling result and the scaling relations. Note that as the stellar density $M/R^3$ is more of less fixed, the radii and masses follow the same trend in Fig.~\ref{fig:comparison}. All proposed corrections to or revisions of the asteroseismic scaling relations decrease the gap between the scaling relations and stellar models.

\section{Conclusions}
\label{sec:conclusions}
We presented an in-depth analysis of the bright F6 subgiant star \mystar using long-baseline optical interferometry, asteroseismology, high-resolution spectroscopy, and grid-based stellar modelling. All results can be found in Table~\ref{tab:par} and our main findings can be summarised as follows:
\begin{itemize}
\item The radius from pure scaling relations and interferometry shows good agreement, which contradicts the results from \cite{huber2017} where subgiants were found to be systematically underestimated compared to radii computed from \emph{Gaia} DR1. The different findings could be due to the use of two very different \emph{Gaia} data releases, with DR2 being the first data release based entirely on \emph{Gaia} data alone. 
\item Revised scaling relations from \citet{white2011}, \citet{sharma2016}, \citet{sahlholdt2018}, \citet{kallinger2018}, and \citet{bellinger2019} all do improve the agreement between asteroseismology and interferometry.
\item Grid-based stellar modelling of the interferometric temperature, spectroscopic metallicity and the frequency pattern systematically finds solutions with smaller radii and smaller masses than those obtained from interferometry and scaling relations. A mixing length parameter lower than the solar-calibrated value by $\Delta \amlt = 0.2$ was needed to reconcile the results from modelling \mystar to the observables. 3D hydrodynamical simulations of stellar atmospheres support a mixing length parameter lower than solar for a star of this temperature, $\log g$, and metallicity.
\end{itemize}
\citet{lebreton2014} examine how the model input physics affect the stellar age and mass of the main-sequence star HD~52265. 
In the future, we plan a similar study of systematics in stellar modelling for subgiant stars like \mystar in order to quantify the impact of the model physics.

\mystar is only the second subgiant star to have high-quality asteroseismic and interferometric data and this kind of benchmark star with independent empirical estimates of stellar parameters such as radius is valuable in order to understand the shortcomings in our stellar models.
NASA's on-going TESS mission \citep{ricker2014} will almost exclusively detect oscillations in subgiant stars, so lessons learned from benchmark stars such as \mystar will be important to fully explore this new era in space exploration.

\section*{Acknowledgements}
We thank Tom Barclay for being PI on \emph{Kepler} Guest Investigator Program GO40009. We would also like to thank J{\o}rgen Christensen-Dalsgaard, Earl Bellinger, Kuldeep Verma, Kosmas Gazeas, David Soderblom, Thomas Kallinger, Dennis Stello, and the anonymous referee for helpful comments and suggestions, which significantly contributed to improving the quality of this paper.

Funding for the Stellar Astrophysics Centre is provided by The Danish National Research Foundation (Grant agreement no.: DNRF106). V.S.A. and T.R.W. acknowledge support from the Villum Foundation (Research grant 10118). V.S.A. acknowledges support from the Independent Research Fund Denmark (Research grant 7027-00096B). T.R.W. acknowledges the support of the Australian Research Council (grant DP150100250). M.N.L. acknowledges the support of The Danish Council for Independent Research | Natural Science (Grant DFF-4181-00415).
D.H. acknowledges support by the National Aeronautics and Space Administration under Grant NNX14AB92G issued through the Kepler Participating Scientist Program and support by the National Science Foundation (AST-1717000).

We acknowledge the {\it Kepler} Science Team and all those who have contributed to the {\it Kepler} mission. Funding for the {\it Kepler} mission is provided by the NASA Science Mission directorate. 
This work is based upon observations obtained from the Georgia State University Center for High Angular Resolution Astronomy Array at Mount Wilson Observatory. The CHARA Array is supported by the National Science Foundation under Grant nos.\@ AST-1211929 and AST-1411654. Institutional support has been provided from the GSU College of Arts and Sciences and the GSU Office of the Vice President for Research and Economic Development. This work also includes observations made with the Hertzsprung SONG telescope operated at the Spanish Observatorio del Teide on the island of Tenerife by the Aarhus and Copenhagen Universities and by the Instituto de Astrof\'isica de Canarias. This research has made use of the SIMBAD data base, operated at CDS, Strasbourg, France. This work has made use of data from the European Space Agency (ESA) mission {\it Gaia}  (\url{http://www.cosmos.esa.int/gaia}), processed by the {\it Gaia} Data Processing and Analysis Consortium (DPAC, \url{http://www.cosmos.esa.int/web/gaia/dpac/consortium}). Funding for the DPAC has been provided by national institutions, in particular, the institutions participating in the {\it Gaia} Multilateral Agreement.



\bibliographystyle{mnras}
\bibliography{hr7322} 

\appendix
\begin{figure*}
	\includegraphics[width=\textwidth]{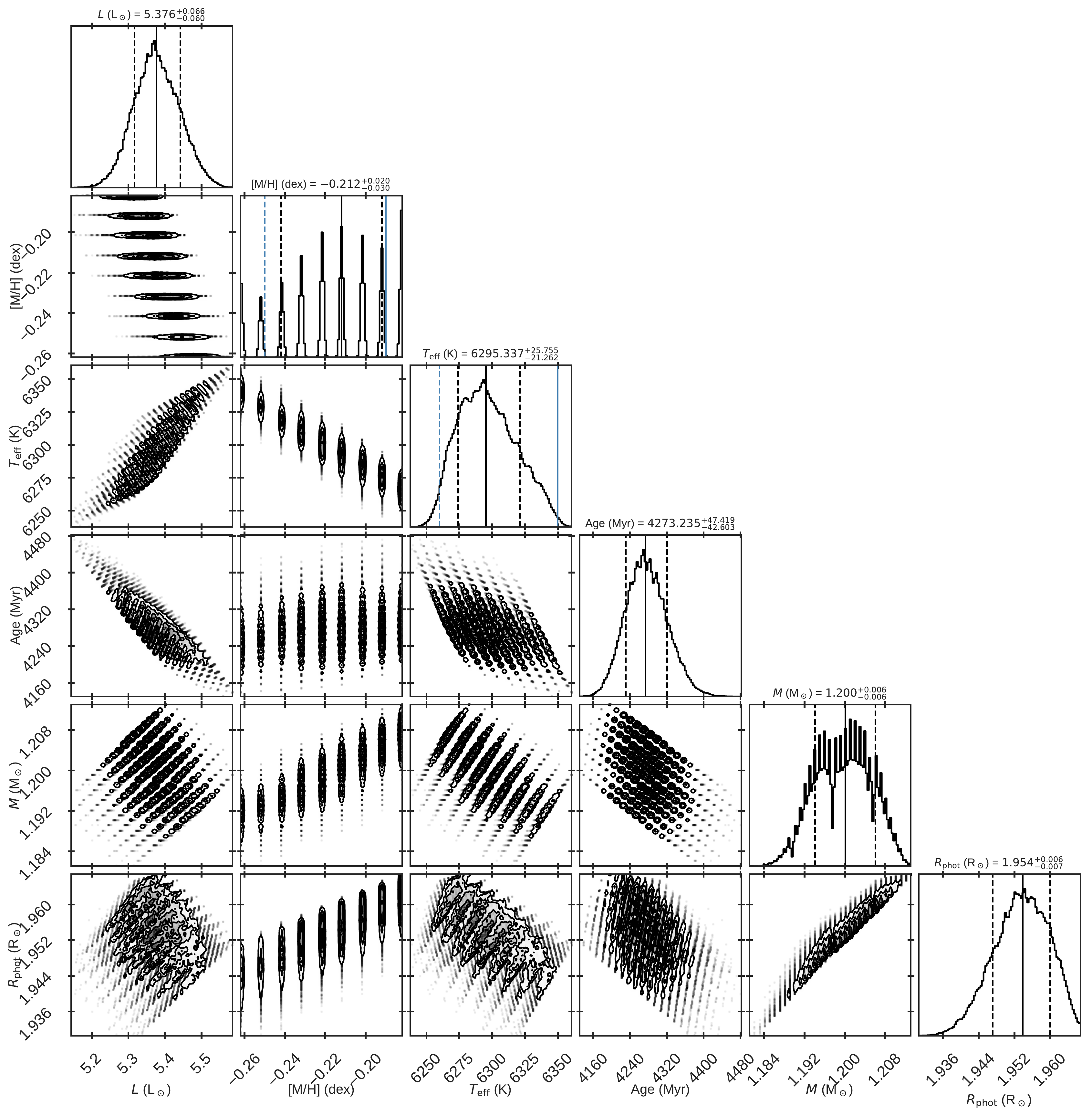}
	\caption{The probability density functions for the refined \emph{lowmlt} grid. The blue solid lines in effective temperature and metallicity show the observed values, and the dashed blue lines indicate the $1\sigma$ uncertainties of the observations.}
	\label{fig:corner}
	\end{figure*}


\bsp	
\label{lastpage}
\end{document}